# Novel Framework for Hidden Data in the Image Page within Executable File Using Computation between Advanced Encryption Standard and Distortion Techniques

*A.W. Naji\*, Shihab A. Hameed, , B.B.Zaidan\*\*, Wadji F. Al-Khateeb, Othman O. Khalifa, A.A.Zaidan and Teddy S. Gunawan,*

Department of Electrical and Computer Engineering, Faculty of Engineering,
International Islamic University Malaysia, P.O. Box 10, 50728 Kuala Lumpur, Malaysia
\* ahmed@iiu.edu.my, \*\* bilal_bahaa@hotmail.com

**ABSTRACT----- The hurried development of multimedia and internet allows for wide distribution of digital media data. It becomes much easier to edit, modify and duplicate digital information. In additional, digital document is also easy to copy and distribute, therefore it may face many threats. It became necessary to find an appropriate protection due to the significance, accuracy and sensitivity of the information. Furthermore, there is no formal method to be followed to discover a hidden data. In this paper, a new information hiding framework is presented.The proposed framework aim is implementation of framework computation between advance encryption standard (AES) and distortion technique (DT) which embeds information in image page within executable file (EXE file) to find a secure solution to cover file without change the size of cover file. The framework includes two main functions; first is the hiding of the information in the image page of EXE file, through the execution of four process (specify the cover file, specify the information file, encryption of the information, and hiding the information) and the second function is the extraction of the existence information through three process (specify the stego file, extract the information, and decryption of the information).**

**Keyword--(Image Pages within Portable Executable File Cryptography, Advance Encryption Standard, Steganography, Distortion Technique).**

## I. INTRODUCTION

Nowadays, protection framework can be classified into more specific as hiding information (Steganography) or encryption information (Cryptography) or a combination between them. Cryptography is the practice of 'scrambling' messages so that even if detected, they are very difficult to decipher. The purpose of Steganography is to conceal the message such that the very existence of the hidden is 'camouflaged'.However, the two techniques are not mutually exclusive. Steganography and Cryptography are in fact complementary techniques [1],[2]. No matter how strong algorithm, if an encrypted message is discovered, it will be subject to cryptanalysis. Likewise, no matter how well concealed a message is, it is always possible that it will be discovered [1]. By combining Steganography with Cryptography we can conceal the existence of an encrypted message. In doing this, we make it far less likely that an encrypted message will be found [2],[3]. Also, if a message concealed through Steganography is discovered, the discoverer is still faced with the formidable task of deciphering it. Also the strength of the combination between hiding and encryption science is due to the non-existence of standard algorithms to be used in (hiding and encryption) secret messages. Also there is randomness in hiding methods such as combining several media (covers) with different methods to pass a secret message. Furthermore, there is no formal method to be followed to discover a hidden data [1],[2],[3].

## II. IMAGE PAGE WITHIN PORTABLE EXECUTABLE FILE

The proposed framework uses a portable executable file as a cover to embed an executable program as an example for the proposed framework. This section is divided into three parts [3],[4],[5]. Firstly concepts related with PE,the addition of the Microsoft® windows NT™ operating system to the family of windows™ operating systems brought many changes to the development environment and more than a few changes to applications themselves. One of the more significant changes is the introduction of the Portable Executable (PE) file format. The name "Portable Executable" refers to the fact that the format is not architecture specific [6].In other words, the term "Portable Executable" was chosen because the intent was to have a common file format for all versions of Windows, on all supported CPUs [5].The PE files formats drawn primarily from the Common Object File Format (COFF) specification that is common to UNIX® operating systems. Yet, to remain compatible with previous versions of the MS-DOS® and windows operating systems, the PE file format also retains the old familiar MZ header from MS-DOS [6].The PE file format for Windows NT introduced a completely new structure to developers familiar with the windows and MS-DOS environments. Yet developers familiar with the UNIX environment will find that the PE file format is similar to, if not based on, the COFF specification [6].The entire format consists of an MS-DOS MZ header, followed by a real-mode stub program, the PE file signature, the PE file header, the PE optional header, all of the section headers, and finally, all of the section bodies [4].The secondly part are techniques related with PE ,before looking inside the PE file, we should know special techniques some of which are [6], general view of PE files sections, a PE file section represents code or data of some sort. While code is just code, there are multiple types of data. Besides read/write program data (such as global variables), other types of data in sections include application program interface (API) import and export tables, resources, and relocations. Each section has its own set of in-memory attributes, including whether the section contains code, whether it's read-only or read/write, and whether the data in the section is shared between all processes using the executable file. Sections have two alignment values, one within the desk file and the other in memory [5]. The PE file header specifies both of these values, which can differ. Each section starts at an offset that's some multiple of the alignment value. For instance, in the PE file, a typical alignment would be 0x200. Thus, every section begins at a file offset that's a multiple of 0x200.Once mapped into memory, sections always start on at least a page boundary. That is, when a PE section is mapped into memory, the first byte of each section corresponds to a memory page. On x86 CPUs, pages are 4KB aligned, while on the Intel Architecture IA-64, they're 8KB aligned. Relative Virtual Addresses (RVA), in an executable file, there are many places where an in-memory address needs to be specified. For instance, the address of a global variable is needed when referencing it. PE files can load just about anywhere in the process address space [7]. While they do have a preferred load address, you can't rely on the executable file actually loading there. For this reason, it's important to have some way of specifying addresses that are independent of where the executable file loads. To avoid having hard coded memory addresses in PE files, RVAs are used. An RVA is simply an offset in memory, relative to where the PE file was loaded. For instance, consider an .EXE file loaded at address 0x400000, with its code section at address 0x401000. The RVA of the code section would be:

**(Target address) 0x401000 – (load address) 0x400000 = (RAV) (1)**

To convert an RVA to an actual address, simply reverse the process: add the RVA to the actual load address to find the actual memory address. Incidentally, the actual memory address is called a Virtual Address (VA) in PE parlance [7]. Another way to think of a VA is that it's an RVA with the preferred load address added in. Importing Functions, when we use code or data from another DLL, we're importing it. When any PE files loads, one of the jobs of the windows loader is to locate all the imported functions and data and make those addressees available to the file being loaded. The thirdly part is The PE file layout is shown in Figure 1. There is image page in PE file layout [7], and this image page suggested to hide a watermark.The size of the image page is different from one file to another [7].The most important reason behind the idea of this framework is that the programmers always need to create a back door for all of their developed applications, as a solution to many problems such that forgetting the password. This idea leads the customers to feel that all programmers have the ability to hack their framework any time. At the end of this discussion all customers always are used to employ trusted programmers to build their own application. Programmers want their application to be safe any where without the need to build ethic relations with their customers. In this framework a solution is suggested for this problem [6],[8]. The solution is to hide the password in the executable file of the same framework and then other application to be retracted by the customer himself. Steganography needs to know all files format to find a way for hiding information in those files. This technique is difficult because there are always large numbers of the file format and some of them have no way to hide information in them [8].

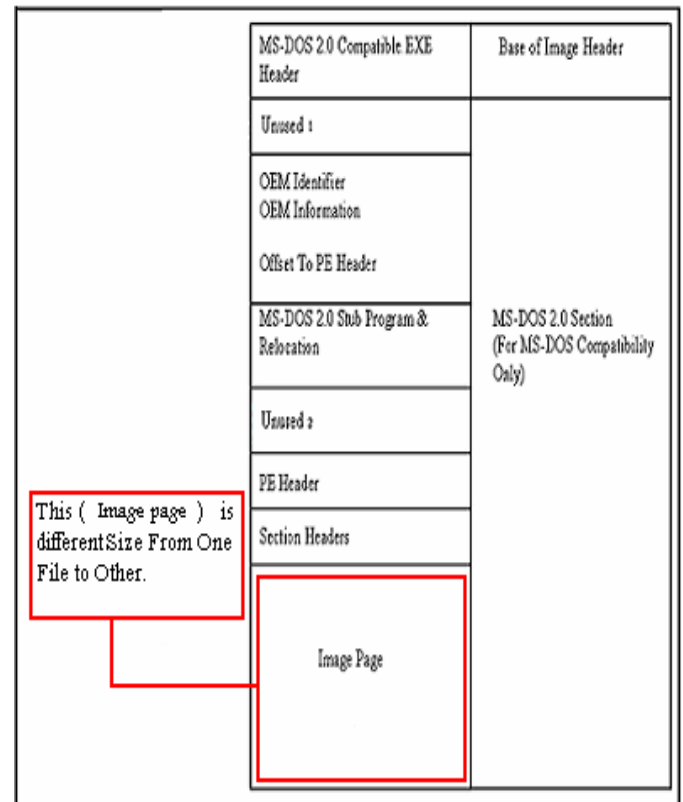

Figure 1.Typical 32-bit Portable .EXE File Layout

## III. CRYPTOGRAPHY

### A. Block Cipher

In cryptography, a block cipher is a symmetric key cipher which operates on fixed-length groups of bits, termed blocks, with an unvarying transformation. When encrypting, a block cipher might take a (for example) 128-bit block of plaintext as input, and outputs a corresponding 128-bit block of cipher text. The exact transformation is controlled using a second input — the secret key. Decryption is similar: the decryption algorithm takes, in this example, a 128-bit block of cipher text together with the secret key, and yields the original 128-bit block of plaintext. To encrypt messages longer than the block size (128 bits in the above example), a mode of operation is used. Block ciphers can be contrasted with stream ciphers; a stream cipher operates on individual digits one at a time and the transformation varies during the encryption. The distinction between the two types is not always clear-cut: a block cipher, when used in certain modes of operation, acts effectively as a stream cipher as shown in Figure 2 [8].

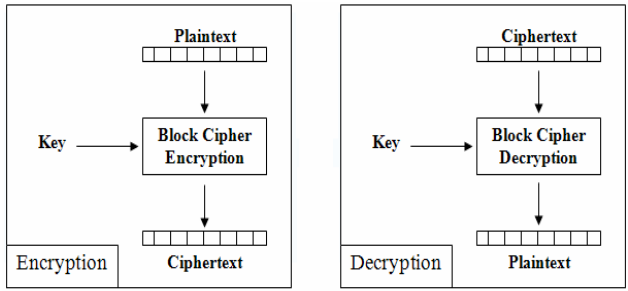

Figure 2. Encryption and Decryption

An early and highly influential block cipher design is the Data Encryption Standard (DES). The (**DES**) is a cipher (a method for encrypting information) selected as an official Federal Information Processing Standard (FIPS) for the United States in 1976, and which has subsequently enjoyed widespread use internationally. The algorithm was initially controversial, with classified design elements, a relatively short key length, and suspicions about a National Security Agency (NSA) backdoor. DES consequently came under intense academic scrutiny, and motivated the modern understanding of block ciphers and their cryptanalysis. DES is now considered to be insecure for many applications. This is chiefly due to the 56-bit key size being too small; DES keys have been broken in less than 24 hours. There are also some analytical results which demonstrate theoretical weaknesses in the cipher, although they are infeasible to mount in practice. The algorithm is believed to be practically secure in the form of Triple DES, although there are theoretical attacks[1][8]. In recent years, the cipher has been superseded by the Advanced Encryption Standard (AES).

*B.  Advanced Encryption Standard*

Advance Encryption Standard (AES) and Triple DES (TDES or 3DES) are commonly used block ciphers. Whether you choose AES or 3DES depend on your needs. In this section it would like to highlight their differences in terms of security and performance [3].Since 3DES is based on DES algorithm, it will talk about DES first. DES was developed in 1977 and it was carefully designed to work better in hardware than software. DES performs lots of bit manipulation in substitution and permutation boxes in each of 16 rounds. For example, switching bit 30 with 16 is much simpler in hardware than software. DES encrypts data in 64 bit block size and uses effectively a 56 bit key. 56 bit key space amounts to approximately 72 quadrillion possibilities. Even though it seems large but according to today's computing power it is not sufficient and vulnerable to brute force attack. Therefore, DES could not keep up with advancement in technology and it is no longer appropriate for security. Because DES was widely used at that time, the quick solution was to introduce 3DES which is secure enough for most purposes today.3DES is a construction of applying DES three times in sequence. 3DES with three different keys (K1, K2 and K3) has effective key length is 168 bits (The use of three distinct key is recommended of 3DES.). Another variation is called two-key (K1 and K3 is same) 3DES reduces the effective key size to 112 bits which is less secure. Two-key 3DES is widely used in electronic payments industry. 3DES takes three times as much CPU power than compare with its predecessor which is significant performance hit. AES outperforms 3DES both in software and in hardware [8]. The Rijndael algorithm has been selected as the Advance Encryption Standard (AES) to replace 3DES. AES is modified version of Rijndael algorithm. Advance Encryption Standard evaluation criteria among others was (Seleborg, 2004):
• Security
• Software & Hardware performance
• Suitability in restricted-space environments
• Resistance to power analysis and other implementation attacks.

Rijndael was submitted by Joan Daemen and Vincent Rijmen. When considered together Rijndael combination of security, performance, efficiency, implement ability, and flexibility made it an appropriate selection for the AES.By design AES is faster in software and works efficiently in hardware. It works fast even on small devices such as smart phones; smart cards etc.AES provides more security due to larger block size and longer keys.AES uses 128 bit fixed block size and works with 128, 192 and 256 bit keys.Rigndael algorithm in general is flexible enough to work with key and block size of any multiple of 32 bit with minimum of128 bits and maximum of 256 bits.AES is replacement for 3DES according to NIST both ciphers will coexist until the year2030 allowing for gradual transition to AES.Even though AES has theoretical advantage over 3DES for speed and efficiency in some hardware implementation 3DES may be faster where support for 3DES is mature [1][2][5].

## IV.  STEGANOGRAPHY

*A.  General Steganography Framework*

A general Steganography framework is shown in Figure 3. It is assumed that the sender wishes to send via Steganographic transmission, a message to a receiver. The sender starts with a cover message, which is an input to the stego-system, in which the embedded message will be hidden. The hidden message is called the embedded message. A Steganographic algorithm combines the cover massage with the embedded message, which is something to be hidden in the cover The algorithm may, or may not, use a Steganographic key (stego key), which is additional secret data that may be needed in the hidden process. The same key (or related one) is usually needed to extract the embedded massage again. The output of the Steganographic algorithm is the stego message. The cover massage and stego message must be of the same data type, but the embedded message may be of another data type. The receiver reverses the embedding process to extract the embedded message [4].

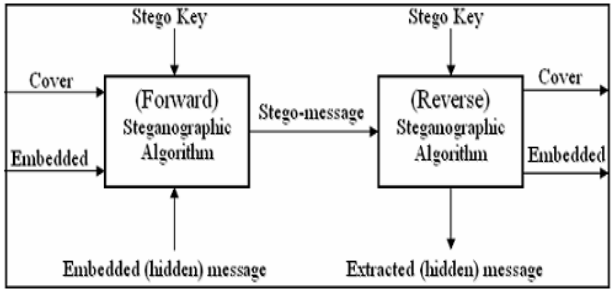

Figure 3: General Steganography Framework

*B. Distortion techniques*

DT is short for Distortion techniques are requires the knowledge of the original cover in the decoding process.The sender applies a sequence of modifications to the cover in order to get a stego-system.A sequence of modification is chosen in such a way that it corresponds to a specific secret message to be transmitted[5].

The receiver measures the difference in the original cover in order to reconstruct the sequence of modification applied by the sender, which corresponds to the secret message. An early approach to hiding information is in text. Most text-based hiding methods are of distortion type (i.e, the arrangement of words or the layout of a docement may reveal information). One technique is by modulating the positions of line and words, which will be detailed in the next subsection.Adding spaces and "invisible" characters to text provides a method to pass hidden information HTML files are good candidates for including image, extra spaces, tabs, and line breaks. The executable file include the image, this point make this cover to be suitable to use for this technique [5].

# V. METHODOLOGY

*A. Framework Concept*

Concept of this framework can be summarized as hiding the password or any information beyond the end of an executable file so there is no function or routine (open-file, read, write, and close-file) in the operating framework to extract it. This operation can be performed in two alternative methods: Building the file handling procedure independently of the operating system file handling routines. In this case we need canceling the existing file handling routines and developing a new function which can perform our need, with the same names.

The advantage of these methods is it doesn't need any additional functions, which can be identified by the analysts. And it can be executed remotely and suitable for networks and the internet applications .The disadvantage of these methods is it needs to be installed (can not be operated remotely). So we choose this concept to implementation in this paper.

*B. Framework Features*

This framework has the following features:

- The hiding operation within image page of EXE file increases the degree of security of hiding technique which is used in the proposed framework because the size of cover file doesn't change, so the attacker can not be attack the information hidden.

- It's very difficult to extract the hidden information it's difficult to find out the information hiding , that is because of three reasons:

  o The information hiding will be encrypted before hiding of the information by AES method; this method very strong, 128-bit key would be in theory being in range of a military budget within 30-40 years. An illustration of the current status for AES is given by the following example, where we assume an attacker with the capability to build or purchase a framework that tries keys at the rate of one billion keys per second. This is at least 1 000 times faster than the fastest personal computer in 2004. Under this assumption, the attacker will need about 10 000 000 000 000 000 000 000 years to try all possible keys for the weakest version.

- The attacker impossible guessing the information hiding inside the EXE file because of couldn't guessing the real size of (EXE file and information hiding).

- The information hiding should be decrypted after retract of the information.

*C. The Proposed Framework Structure.*

To protect the hidden information from retraction the framework encrypts the information by the built-in encryption algorithm provided by the Java. The following Framework algorithm for hiding operation procedure as shown in Figure 4. The following Framework algorithm for Retract operation procedure as shown in Figure 5.

**1. The following algorithm is the hiding operation procedure:**

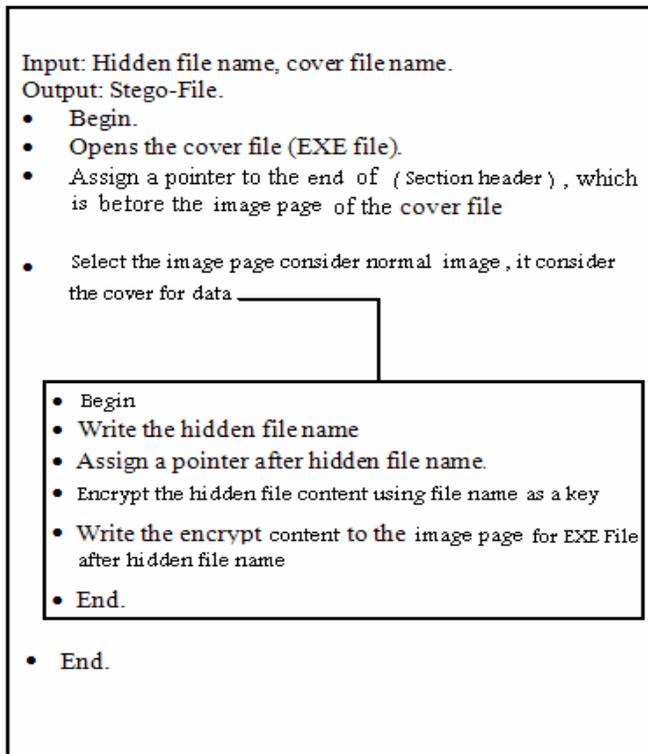

Figure 4. Shows Framework Algorithm for Hiding Operation.

**2. The following algorithm is retraction operation procedure:**

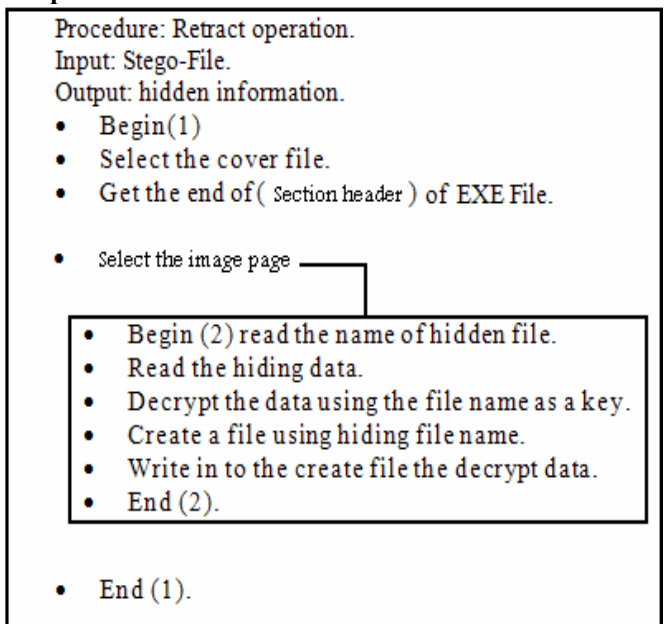

Figure 5. Shows Framework Algorithm for Retract Operation.

## VI. CONCLUSION

One of the important conclusions in implementation of the proposed framework is the solving of the problems that are related to the change size of cover file, so the hiding method makes the relation between the cover and the message dependent without change of cover file and The encryption of the message increases the degree of security of hiding technique which is used in the proposed framework and PE files structure is very complex because they depend on multi headers and addressing, and then insertion of data to PE files without full understanding of their structure may damage them, so the choice is to hide the information beyond the structure of these files , finally The framework has achieved the main goal, makes the relation of the size of the cover file and the size of information dependent without change the size of cover file , so There is no change on the cover file size where you can hide file of image page within portable executable file by Structure on the property of  the EXE file and The proposed framework is implemented by using  Java.

## VII. FUTURE WORK

There are many suggestions for improving the proposed framework, the main suggestions are:

- Improvement of the security of hiding framework by adding compression function of the message before the hidden operation.
- Improvement of the security of the proposed framework by changing the encryption methods for other methods such as (MD5, BLOWFISH, DIGITAL SIGNATURE or KEY DISTRIBUTION).

## AUTHORS PROFIL

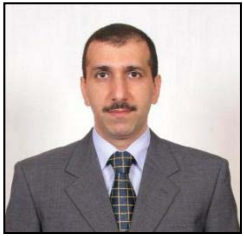

**Dr. Ahmed Wathik Naji -** He obtained his 1st Class Master degree in Computer Engineering from University Putra Malaysia followed by PhD in Communication Engineering also in University Putra Malaysia. He supervised many postgraduate students and led many funded research projects with more than 50 international papers. He has more than 10 years of industrial and educational experience. He is currently Senior Assistant Professor, Department of Electrical and Computer Engineering, International Islamic University Malaya, Kuala Lumpur, Malaysia.

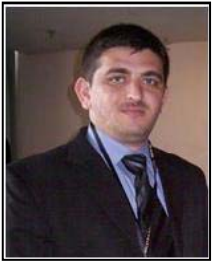

**Aos Alaa Zaidan -** He obtained his 1st Class Bachelor degree in Computer Engineering from university of Technology / Baghdad followed by master in data communication and computer network from University of Malaya. He led or member for many funded research projects and He has published more than 40 papers at various international and national conferences and journals, he has done many projects on Steganography for data hidden through different multimedia carriers image, video, audio, text, and non multimedia carrier unused area within exe.file, Quantum Cryptography and Stego-Analysis systems, currently he is working on the multi module for Steganography. He is PhD candidate on the Department of Computer System & Technology / Faculty of Computer Science and Information Technology/University of Malaya /Kuala Lumpur/Malaysia.

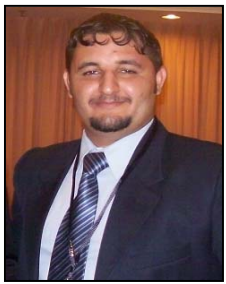

**Bilal Bahaa Zaidan** - he obtained his bachelor degree in Mathematics and Computer Application from Saddam University/Baghdad followed by master from Department of Computer System & Technology Department Faculty of Computer Science and Information Technology/University of Malaya /Kuala Lumpur/Malaysia, He led or member for many funded research projects and He has published more than 40 papers at various international and national conferences and journals. His research interest on Steganography & Cryptography with his group he has published many papers on data hidden through different multimedia carriers such as image, video, audio, text, and non multimedia careers such as unused area within exe.file, he has done projects on Stego-Analysis systems, currently he is working on Quantum Key Distribution QKD and multi module for Steganography, he is PhD candidate on the Department of Computer System & Technology / Faculty of Computer Science and Information Technology/University of Malaya /Kuala Lumpur/Malaysia.

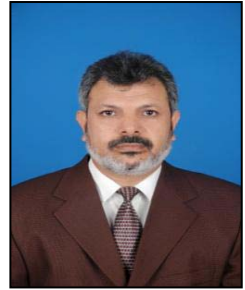

**Dr. Shihab A Hameed -** He obtained his PhD in software Engineering from UKM. He has three decades of industrial and educational experience. His research interest is mainly in the software engineering, software quality, surveillance and monitoring systems, health care and medication. He supervised numerous funded projects and has published more than 60 papers at various international and national conferences and journals. He is currently Senior Assistant Professor, Department of Electrical and Computer Engineering, International Islamic University Malaysia. Malaya, Kuala Lumpur.

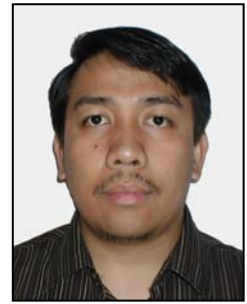

**Dr. Teddy Surya Gunawan** - He received his B.Eng degree in Electrical Engineering with cum laude award from Institut Teknologi Bandung (ITB), Indonesia in 1998. He obtained his M.Eng degree in 2001 from the School of Computer Engineering at Nanyang Technological University, Singapore, and PhD degree in 2007 from the School of Electrical Engineering and Telecommunications, The University of New South Wales, Australia. His research interests are in speech and image processing, biomedical processing, image processing, and parallel processing. He is currently Assistant Professor and Academic Advisor at Department of Electrical and Computer Engineering, International Islamic University Malaysia. He is currently Senior Assistant Professor at Department of Electrical and Computer Engineering, International Islamic University Malaysia.

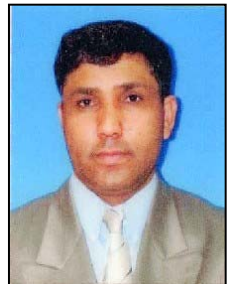

**Othman O. Khalifa** received his Bachelor's degree in Electronic Engineering from the Garyounis University, Libya in 1986. He obtained his Master degree in Electronics Science Engineering and PhD in Digital Image Processing from Newcastle University, UK in 1996 and 2000 respectively. He worked in industrial for eight years and he is currently an Professor and Head of the department of Electrical and Computer Engineering, International Islamic University Malaysia. His area of research interest is Communication Systems, Information theory and Coding, Digital image / video processing, coding and Compression, Wavelets, Fractal and Pattern Recognition. He published more than 130 papers in international journals and Conferences. He is SIEEE member, IEEE computer, Image processing and Communication Society member.

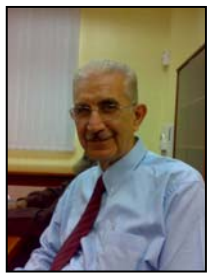

**Dr. Wajdi Fawzi Al-Khateeb**- He received his PhD from the International Islamic University, Malaysia and his MSc from the Technical University of Berlin, Germany. His research interest is mainly in the Reliability Engineering, Fault Tolerant Systems, QoS Networking, Microwave Radio Links. He is currently an assistant Professor in the department of Electrical and Computer Engineering, International Islamic University Malaysia